\documentclass[aps,prd,twocolumn,showpacs,superscriptaddress,nofootinbib,floatfix,showkeys,10pt]{revtex4-2}

\usepackage[colorlinks=true, linkcolor=blue, urlcolor=blue, citecolor=blue]{hyperref}
\usepackage{bm}
\usepackage{times}
\usepackage{braket}
\usepackage{amsfonts,amssymb,stmaryrd,latexsym,amsmath}
\usepackage[usenames,dvipsnames]{color}
\usepackage{slashed}
\usepackage{hyperref}
\usepackage{subfigure}
\usepackage{graphicx}
\usepackage{orcidlink}
\usepackage{multirow}
\usepackage{appendix}
\usepackage{dashrule}
\usepackage{float}
\allowdisplaybreaks[1]
\usepackage{verbatim}
\usepackage{overpic}
\newcommand{\sysu}{\affiliation{School of Physics, Sun Yat-sen University, Guangzhou 510275, China}}

\begin{document}

\title{Coupled-channel study of the three-body $DDK$ and $D^{*}D^{*}K$}

\author{Hai-Peng Xie}\sysu

\author{Si-Yi Chen}\sysu


\author{Ning Li\,\orcidlink{0000-0003-2987-2809}}\email{lining59@mail.sysu.edu.cn}\sysu
\author{Wei Chen\,\orcidlink{0000-0002-8044-5493}}\email{chenwei29@mail.sysu.edu.cn}
\sysu
\affiliation{Southern Center for Nuclear-Science Theory (SCNT), Institute of Modern Physics, 
Chinese Academy of Sciences, Huizhou 516000, Guangdong Province, China}

\begin{abstract}

We investigate the three-body $DDK$ system with quantum numbers
 $I(J^P) = \frac{1}{2}(0^-)$ within a coupled-channel framework that 
 incorporates both $DDK$ and $D^{*}D^{*}K$ configurations. 
 The $D^{(*)}D^{(*)}$ interactions are described using the 
 one-boson-exchange model constrained by heavy-quark symmetry 
 and fitted to the pole positions of $X(3872)$, $T_{cc}^+$, and $Z_c(3900)$. 
 The $D^{(*)}K$ interaction is from the chiral effective theory, motivated by 
 the molecular interpretation of $D_{s0}^*(2317)$, and is further constrained 
 by lattice-QCD results for the $DK$ scattering lengths.
The resulting three-body problem is solved using the Gaussian expansion 
method, while the complex scaling method is employed to search for possible 
resonant states. We find that coupled-channel effects from $D^*D^*K$ are negligible, 
and the $DDK$ system supports a deeply bound state across a wide range of parameters. 
Depending on the long-range behavior of the $DK$ interaction, an additional shallow 
state may emerge near the particle–dimer ($D$-$DK$) threshold. The deeply bound state exhibits a compact three-body structure, whereas 
the shallow state displays characteristic features of a three-body halo configuration. 
No clear resonance poles are identified within the explored parameter region. 
Similar results are obtained for the $D^*D^*K$ system. These findings may provide 
new insight into few-body dynamics in systems involving charmed mesons and kaons.
\end{abstract}

\maketitle

\section{\label{sec:level1}Introduction}

The spectroscopy of hadrons containing heavy quarks has undergone rapid 
development in recent decades. Numerous unconventional states have been 
observed that cannot be easily accommodated within the conventional 
quark–antiquark or three-quark classification scheme. These states are 
commonly referred to as exotic hadrons. The proposed models to 
interpret them include tetraquarks, pentaquarks, hadronic molecules, etc. 
Their existence provides valuable insight into the nonperturbative dynamics 
of quantum chromodynamics (QCD), see reviews~~\cite{Chen:2016qju,Guo:2017jvc,Chen:2022asf,Liu:2019zoy,Chen:2016spr,Liu:2024uxn,Guo:2019twa,Wang:2025sic,Brambilla:2019esw,Meng:2022ozq,Esposito:2016noz,Lebed:2016hpi} .

One particularly important class of exotic candidates consists of 
hadronic molecules formed by two or more hadrons interacting through 
residual strong forces. In the heavy-quark sector, heavy-quark spin 
symmetry (HQSS) implies close dynamical connections among systems containing 
heavy mesons such as $D$ and $D^*$~\cite{Yan:1992gz,Casalbuoni:1996pg,Georgi:1990um,Manohar:2000dt}. 
The systems composed of charmed mesons provide a natural laboratory for exploring molecular 
configurations. From an experimental perspective, many open/hidden-charm exotic hadrons have been observed  
since the discovery of $X(3872)$ by the Belle Collaboration 
in 2003 \cite{Belle:2003nnu}. For example, the charged charmonium-like state $Z_c(3900)$ 
was observed by the BESIII Collaboration~\cite{Ablikim:2013mio} and Belle Collaboration~\cite{Liu:2013dau}
in  2013, the doubly charmed 
tetraquark state $T_{cc}^+$ was observed by the LHCb Collaboration in 2021 
\cite{Aaij:2021vvq,Aaij:2021ivw}, and the hidden-charm 
pentaquark family $P_c$s were reported by the LHCb Collaboration in 2015 and 2019 
\cite{Aaij:2015tga,Aaij:2019vzc}, refer to ~\cite{Swanson:2006st,Voloshin:2007dx,Brambilla:2010cs,Brambilla:2014jmp,Hosaka:2016pey,Chen:2016qju,Esposito:2016noz,Lebed:2016hpi,Ali:2017jda,Guo:2017jvc,Olsen:2017bmm,Liu:2019zoy,Brambilla:2019esw,Meng:2022ozq,Chen:2022asf,Liu:2024uxn,Wang:2025sic} for comprehensive reviews.

Among the exotic hadrons observed experimentally so far, $D_{s0}^*(2317)$, 
observed by the BABAR Collaboration \cite{BaBar:2003oey} 
and later confirmed by the CLEO \cite{CLEO:2003ggt} and Belle 
\cite{Belle:2003guh} Collaborations is particularly interesting. The unexpectedly low mass 
of $D_{s0}^*(2317)$ has been interpreted as the evidence 
for a $DK$ hadronic molecule~\cite{Kolomeitsev:2003ac,Guo:2006fu,Barnes:2003dj,vanBeveren:2003ds,Hofmann:2004opencharm,Gamermann:2007dynamic}. 
Its heavy-spin partner, $D_{s1}$(2460), was observed by several experimental 
collaborations~\cite{2460Babar,CLEO:2003ggt,Belle:2003guh,2460LHCb}, and exhibits similarly anomalous properties of $D_{s0}^*(2317)$, making it a candidate for a $D^*K$ molecular state~\cite{Faessler:2007us,Albaladejo:2018ds1,Xiao:2016hoa,Fu:2022ds1} or a multiquark state~\cite{Tan:2021bvl}.  
Furthermore, the $DK$ interaction has also been investigated by lattice-QCD calculations, 
in which the $DK$ scattering length and the 
properties of $D_{s0}^*(2317)$ were extracted 
\cite{Liu:2012zya,Mohler:2013rwa,Lattice_scat_length,DKlattice1,DKlattice2,DKlattice3,Cheung:2020mql}. These studies indicate 
that the $DK$ interaction in the isoscalar ($I = 0$) channel is  
strongly attractive.
More analyses further sharpened this near-threshold picture from coupled-channel perspectives~\cite{Lang:2014ds,Yang:2022nearthreshold}.

Motivated by the strong attraction in the $DK$ channel, three-body systems composed 
of two charmed mesons and one kaon, such as $DDK$, have attracted sustained attention. 
Early studies suggested that such systems may 
support exotic bound states driven by the underlying two-body dynamics. In particular, 
the $DD^{*}K$ system was found within the Born--Oppenheimer framework to admit 
a bound-state solution associated with a delocalized pion-bond mechanism~\cite{Ma:2017ery}. 
A Gaussian expansion method analysis of the $DK$, $DDK$, and $DDDK$ systems 
showed that the strong $DK$ attraction can generate a $DDK$ bound state with 
$I(J^{P})=\frac{1}{2}(0^-)$ and may even support a $DDDK$ four-body state~\cite{Wu:2019vsy}. 
Besides, within a coupled-channel approach, including channels $DD_s\eta$ and $DD_s\pi$, 
the $DDK$ system was found to support a bound state~\cite{MartinezTorres:2018ziv}.
Related few-body studies further emphasized that these systems provide a useful laboratory 
for understanding how effective two-body interactions give rise to nontrivial three-body 
dynamics~\cite{MartinezTorres:2020hus}. More recently, the discussion has been extended
to nearby analog systems: the $D\bar{D}K$ system was shown in the fixed-center 
approximation to develop a bound or resonant structure around $4180$~MeV~\cite{Wei:2022jgc}. 
The lattice effective field theory study indicated that the $DD^{*}K$ system remains bound even 
in the presence of repulsive three-body forces~\cite{Zhang:2024yfj}, and subsequent analyses 
showed that stronger three-body repulsion enlarges the spatial size of the $DD^{(*)}K$ molecule,
eventually leading to a configuration resembling a $D^{(*)}K$ cluster plus a distant $D$ meson, 
whereas the $DDK$ system tends to preserve an isosceles-triangle geometry~\cite{Pan:2025xvq}. 
In addition, the recently predicted $J^{PC}=0^{--}$ $\bar D_sDK$ bound state suggests that such 
three-body molecules may provide further insight into the nature of the $D_{s0}^*(2317)$ and 
enrich the spectrum of exotic multihadron states~\cite{Wu:2025fzx}. Refer to \cite{Liu:2024uxn} for a recent 
review on the three-body $DDK$-like systems. Furthermore, $BK$, as the heavy-flavor partner of $DK$, 
has also been the subject of several dedicated investigations~\cite{BK1,BK2,BK3,BK4,Fu:2022ds1}.

Despite previous investigations, the structure of the $DDK$ system 
remains incompletely understood. In particular, the role of coupled-channel 
dynamics involving the $D^*D^*K$ configuration and the sensitivity of 
the three-body spectrum to the underlying two-body interactions require 
further clarification. The strong attraction in the $DK$ subsystem may 
generate nontrivial three-body dynamics, potentially leading to bound 
states or even resonances.

In this work, we perform a coupled-channel study of the $DDK$ system  incorporating both $DDK$ and 
$D^*D^*K$ configurations  with 
quantum numbers $I(J^P)=\frac{1}{2}(0^-)$. The two-body interactions are 
constructed using a hybrid approach that combines the one-boson-exchange (OBE)
model for the $D^{(*)}D^{(*)}$ sector with chiral-motivated interactions 
for the $D^{(*)}K$ system. For the $D^{(*)}K$ interaction, we adopt the framework of 
Ref.~\cite{Guo:2006fu}. In this model, once two cutoff parameters and a
short-range repulsive coupling are specified, the remaining running coupling is
fixed by reproducing the $D_{s0}^*(2317)$ pole position. 
To further reduce model ambiguity, we add one more constraint from the 
lattice-QCD results of the $DK$ scattering length~\cite{Liu:2012zya,Mohler:2013rwa,Lattice_scat_length}.

Regarding the $D^{(*)}D^{(*)}$ interaction, we adopt the results from the OBE 
model, which is widely applied to systems containing charmed mesons.
In OBE model, a form factor with momentum cutoff is generally applied to 
regularize the interaction to remove the singularity at very short range but 
keep the long-range structure unchanged. 
By varying the cutoff, one performs different regularization to the short-range interaction, 
and thus different final interactions are obtained. 
Complementary lattice-QCD and coupled-channel studies of the doubly charmed sector  
provide useful constraints on the near-threshold $DD^{*}$ dynamics~\cite{Chen:2022tcc,Padmanath:2022tcc,Du:2022tcc,Lyu:2023tcc,Whyte:2025ddstar}. 
In Refs.~\cite{Zhu:2024hgm,Xu:2025mhc,Chen:2026bca},  a new 
approach was proposed to determine the $D^{(*)}D^{(*)}$ interaction, with which 
the coupling constants are redetermined by fitting the 
experimental information of $Z_c(3900), X(3872)$, and $T_{cc}^+$. 
Since with this new approach, the interaction is determined directly from the experimental data, 
but not indirectly from other processes as the generally used OBE potential,
this new interaction is believed to less model-dependent than the conventional one. 

In the present work, the three-body Schr\"odinger equation is solved using the Gaussian expansion 
method (GEM)~\cite{Hiyama:2003cu}, which has proven highly efficient for few-body systems. 
Possible resonance structures are searched for using the complex scaling 
method (CSM)~\cite{Aguilar:1971ve,Balslev:1971vb}. The main goals of this work include 
\begin{itemize}
\item to determine whether the $DDK$ system supports bound or resonant states 
applying the two-body interactions constrained by the experimental data and lattice-QCD calculations,
\item to analyze the internal structures of the resulting states,
\item to clarify the role of coupled-channel effects from $D^*D^*K$.
\end{itemize}

The paper is organized as follows. In Sec.~\ref{sec:level2}, we introduce the two-body 
interactions used in the calculation. In Sec.~\ref{sec:level3}, we present the three-body 
formalism and numerical methods. Results are discussed in Sec.~\ref{sec:level4}, and 
a summary is given in Sec.~\ref{sec:level5}.

\section{\label{sec:level2}Two-body interactions}

\begin{figure}[htbp]
  \centering
  \begin{overpic}[width=0.5\textwidth]{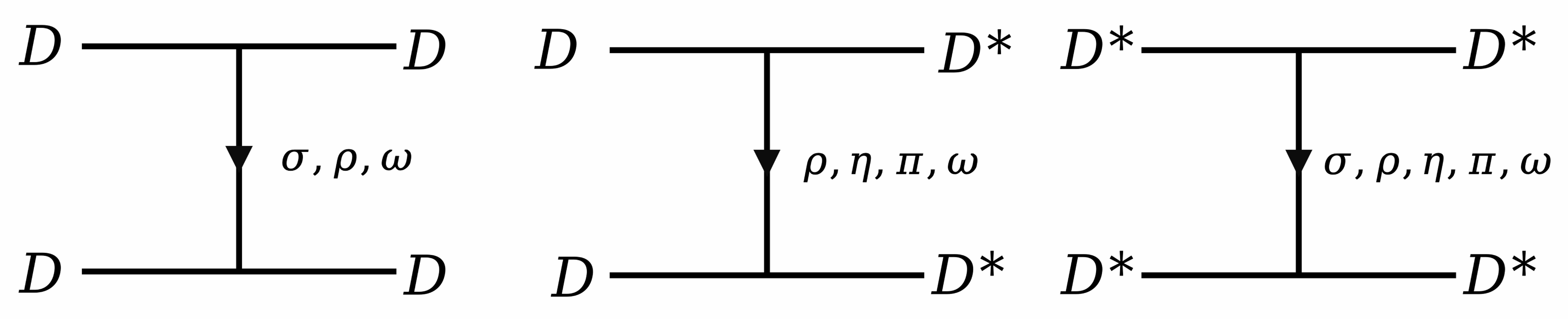}
    \put(13,-2){\small (a)}
    \put(48,-2){\small (b)}
    \put(82,-2){\small (c)}
  \end{overpic}
  \caption{Tree-level feynman diagrams used to derive the OBE potentials: (a) $DD \to DD$, (b) $DD \to D^{*}D^{*}$, and (c) $ D^{*}D^{*} \to D^*D^*$.}
  \label{fig:feynman}
\end{figure}

In the present work, we use the two-body $D^{(*)}K$ and 
$D^{(*)}D^{(*)}$ interactions to rebuild the 
interactions for the coupled channels $DDK$ and $D^{*}D^{*}K$. 
In Fig.~\ref{fig:feynman}, the schematic tree-level Feynman diagrams are presented. 

\subsection{$D^{(*)}D^{(*)}$ interaction}

We derive the $D^{(*)}D^{(*)}$ interaction within the OBE model. 
We start from the Lagrangians in heavy quark effective field theory, 
which are constructed based on heavy-quark
symmetry (including heavy-quark flavor symmetry and heavy-quark spin symmetry), 
chiral symmetry, and SU(2) isospin symmetry.
Under heavy-quark spin symmetry, the pseudoscalar $D$ meson and vector 
$D^{*}$ meson assemble  into a  superfield $\mathcal{H}$ \cite{Casalbuoni:1996pg,OBE}
\begin{align}
    \mathcal{H} = \frac{1 + \slashed{v}}{2} 
    \left(
        \mathit{P}_{\mu}^{*}{\gamma}^{\mu} - \mathit{P}{\gamma}_{5}
    \right),
\end{align}
in which $P = \left(D^{0}, D^{+}\right)$ and 
$P_\mu^{*} = \left(D^{*0}, D^{*+}\right)$. $v$ is the velocity of the charmed meson $D$ or $D^*$, 
and in the heavy-quark mass limit, $m_Q\to \infty$, the static limit $v^\mu = (1, {\vec 0})$ is adopted.  

The Lagrangian for the charmed meson read
\begin{align}
    \mathcal{L} =    &  i \beta \mathrm{Tr} \left[\mathcal{H} v_\mu \left(\mathcal{V}^\mu - \rho^\mu\right) \bar{\mathcal{H}}\right]
        + i \lambda \mathrm{Tr} \left[\mathcal{H} \sigma_{\mu\nu} F^{\mu\nu} \bar{\mathcal{H}}\right] \nonumber \\ 
        & +  g_s \mathrm{Tr} \left[\mathcal{H} \sigma \bar{\mathcal{H}}\right]
        + i g_a \mathrm{Tr} \left[\mathcal{H} \gamma_\mu \gamma_5 \mathcal{A}^\mu \bar{\mathcal{H}}\right],   \label{lagrangian-1}  
\end{align}
where the conjugate field is defined as  
\begin{eqnarray}
\bar{H} \equiv \gamma^0 H^\dag \gamma^0. 
\end{eqnarray}
The field strength tensor of vector mesons is denoted by $F^{\mu\nu}=\partial^\mu \rho^\nu-\partial^\nu \rho^\mu-[\rho^\mu,\rho^\nu]$, while $\mathcal{V}^\mu$ and $\mathcal{A}^\mu$
are the vector and axial currents respectively, for pseudoscalar mesons,
\begin{align}
    \mathcal{V}^\mu = \frac{1}{2} [\xi^\dagger,\partial_\mu\xi], \:
    \mathcal{A}^\mu = \frac{1}{2} \{\xi^\dagger,\partial_\mu\xi\}, \:
    \xi = \mathrm{exp}(i \mathbb{P} / f_\pi).
\end{align}
The vector and pseudoscalar meson matrices are respective
\begin{align}
    \rho^\mu = \frac{i g_V}{\sqrt{2}}
    \begin{bmatrix}
        \frac{\rho^0 + \omega}{\sqrt{2}} & \rho^+ \\
        \rho^- & \frac{-\rho^0 + \omega}{\sqrt{2}}
    \end{bmatrix}^\mu, \nonumber\\
    \mathbb{P} = 
    \begin{bmatrix}
        \frac{\pi^0}{\sqrt{2}} + \frac{\eta}{\sqrt{6}} & \pi^+ \\
        \pi^- & -\frac{\pi^0}{\sqrt{2}} + \frac{\eta}{\sqrt{6}}
    \end{bmatrix}.
\end{align}

Using the Lagrangians presented in Eq.~\ref{lagrangian-1}, we derive the scattering amplitude by evaluating the Feynman diagrams shown in Fig.~\ref{fig:feynman}. The effective potential between two charmed mesons in momentum space is then obtained via the Breit approximation. The specific expressions are given in 
Table~\ref{tab:two}. The effective potential has the following general form, 
\begin{align}
V({\bm k}) = C_{\rm coupling} \times \mathcal{O}_{r, s} \times \mathcal{O}_{\rm iso}. 
\end{align}

\begin{table*}[t]
\caption{Effective momentum-space potentials for scattering processes $DD\to DD$, $D^{*}D^{*}\to D^*D^*$, 
and $DD \to D^{*}D^{*}$.}
\label{tab:two}
\renewcommand{\arraystretch}{2.5}
\begin{ruledtabular}
\begin{tabular}{lccccc}
$V$ & $C_{\mathrm{coupling}}$ 
    & $\mathcal{O}_{r,s}$ 
    & \multicolumn{3}{c}{$\mathcal{O}_{\mathrm{iso}}$} \\
\cline{4-6}
 &  &  & $DD$ & $D^{*}D^{*}$ & $DD \to D^{*}D^{*}$ \\
\hline
$V_{\sigma}^{\mathrm{S}}$
& $g_s^2$ 
& $-\dfrac{\boldsymbol{a}\!\cdot\!\boldsymbol{b}}{u^2+\boldsymbol{k}^2}$
& $\boldsymbol{I}$ & $\boldsymbol{I}$ & $0$ \\

$V_{\rho}^{\mathrm{V}}$
& $\dfrac{\beta^2 g_V^2}{2}$ 
& $-\dfrac{\boldsymbol{a}\!\cdot\!\boldsymbol{b}}{u^2+\boldsymbol{k}^2}$
& $-\dfrac{\boldsymbol{\tau}_1\!\cdot\!\boldsymbol{\tau}_2}{2}$
& $-\dfrac{\boldsymbol{\tau}_1\!\cdot\!\boldsymbol{\tau}_2}{2}$
& $0$ \\

$V_{\omega}^{\mathrm{V}}$
& $\dfrac{\beta^2 g_V^2}{2}$ 
&  $-\dfrac{\boldsymbol{a}\!\cdot\!\boldsymbol{b}}{u^2+\boldsymbol{k}^2}$
& $-\dfrac{1}{2}\boldsymbol{I}$
& $-\dfrac{1}{2}\boldsymbol{I}$
& $0$ \\

$V_{\pi}^{\mathrm{P}}$
& $\dfrac{g_a^2}{f_{\pi}^2}$ 
& $\dfrac{(\boldsymbol{k}\!\cdot\!\boldsymbol{a})(\boldsymbol{k}\!\cdot\!\boldsymbol{b})}{u^2+\boldsymbol{k}^2}$
& $0$
& $\dfrac{\boldsymbol{\tau}_1\!\cdot\!\boldsymbol{\tau}_2}{2}$
& $\dfrac{\boldsymbol{\tau}_1\!\cdot\!\boldsymbol{\tau}_2}{2}$ \\

$V_{\eta}^{\mathrm{P}}$
& $\dfrac{g_a^2}{f_{\pi}^2}$ 
&  $\dfrac{(\boldsymbol{k}\!\cdot\!\boldsymbol{a})(\boldsymbol{k}\!\cdot\!\boldsymbol{b})}{u^2+\boldsymbol{k}^2}$
& $0$
& $\dfrac{1}{6}\boldsymbol{I}$
& $\dfrac{1}{6}\boldsymbol{I}$ \\

$V_{\rho}^{\mathrm{T}}$
& $2\lambda^2 g_V^2$ 
& $\dfrac{(\boldsymbol{k}\!\cdot\!\boldsymbol{a})(\boldsymbol{k}\!\cdot\!\boldsymbol{b})
-\boldsymbol{k}^2(\boldsymbol{a}\!\cdot\!\boldsymbol{b})}{u^2+\boldsymbol{k}^2}$
& $0$
& $-\dfrac{\boldsymbol{\tau}_1\!\cdot\!\boldsymbol{\tau}_2}{2}$
& $-\dfrac{\boldsymbol{\tau}_1\!\cdot\!\boldsymbol{\tau}_2}{2}$ \\

$V_{\omega}^{\mathrm{T}}$
& $2\lambda^2 g_V^2$ 
& $\dfrac{(\boldsymbol{k}\!\cdot\!\boldsymbol{a})(\boldsymbol{k}\!\cdot\!\boldsymbol{b})
-\boldsymbol{k}^2(\boldsymbol{a}\!\cdot\!\boldsymbol{b})}{u^2+\boldsymbol{k}^2}$
& $0$
& $-\dfrac{1}{2}\boldsymbol{I}$
& $-\dfrac{1}{2}\boldsymbol{I}$ \\
\end{tabular}
\end{ruledtabular}
\begin{flushleft}
\footnotesize
\textit{Notes:} Superscripts $S$ and $P$ denote the effective potentials for $\sigma$-exchange and 
pseudoscalar-exchange, respectively while  $V$ and $T$ means the respective vector- and tensor-type 
potential for the vector meson exchange.  
$\mathcal{O}_{r,s}$ and $\mathcal{O}_{\mathrm{iso}}$ represent spin-related  and isospin operators, respectively. $\boldsymbol{k}=\boldsymbol{p}'-\boldsymbol{p}$ is the momentum of the exchanged light 
meson. For $DD$ systems, $\boldsymbol{a}\!\cdot\!\boldsymbol{b}\to1$, for the transition $DD\to D^*D^*$ system, the substitution
$\boldsymbol{a}\to \boldsymbol{\epsilon_3^\dag}, \boldsymbol{b}\to \boldsymbol{\epsilon_4^\dag}$ is required;
and for $D^*D^*$ system, the substitution $\boldsymbol{a}\to \boldsymbol{\epsilon_1}\!\times\!\boldsymbol{\epsilon_3^\dag}, 
\boldsymbol{b}\to \boldsymbol{\epsilon_2}\!\times\!\boldsymbol{\epsilon_4^\dag}$ is required.
\end{flushleft}
\end{table*}

However, the effective potentials associated with light pseudoscalar meson exchange and vector meson exchange exhibit short-range singularities, where the two charmed mesons overlap strongly and the OBE model becomes unreliable. To address this issue, a form factor with a momentum cutoff is commonly introduced to regularize the high-momentum behavior of the potential. The coordinate-space effective potential $V(r)$ is then derived via Fourier transform:
\begin{align}
V^\Lambda(r) = \int \frac{d{\bf k}}{(2\pi)^3} e^{i{\bf k} \cdot {\bf r}} V(\boldsymbol{k},\mu) \mathcal{F}^2(k^2, \Lambda),
\end{align}
where the monopole form factor adopted in the present work is given by
\begin{eqnarray}
\mathcal{F}(k^2, \Lambda) = \frac{\Lambda^2-\mu^2}{\Lambda^2+\boldsymbol{k}^2}.
\end{eqnarray}
By varying $\Lambda$, different regularization procedures are applied to the short-range potential, 
yielding $\Lambda$-dependent effective potentials.

In Refs.~\cite{Zhu:2024hgm,Xu:2025mhc,Chen:2026bca}, a novel approach was employed to 
determine the effective interaction between two charmed mesons. Specifically, the coupling 
constants were extracted by fitting to the experimental data for $Z_c(3900), X(3872)$, and $T_{cc}^+$.

Following Ref.~\cite{Zhu:2024hgm}, we treat the scalar coupling constant as the sole free 
parameter for the $Z_c$ channel and fix it using the $Z_c$ pole position. This is 
motivated by the expectation that the vector $\rho$-exchange and $\omega$-exchange 
contributions largely cancel each other, such that the $\sigma$-exchange interaction 
dominates in the $Z_c(3900)$ system. The vector coupling constants are determined 
from $X(3872)$ and $T_{cc}^+$. For simplicity, we introduce the following scaling factors:
\begin{align}
    \lambda \to \lambda R_\lambda,\quad \beta \to \beta R_\beta,\quad g_s \to g_s R_s.
\end{align}
The baseline coupling constants and meson masses are summarized in Table \ref{tab:basepara}. 
\begin{table}
    \caption{Baseline coupling constants and the masses of the mesons~\cite{ParticleDataGroup:2022pth}.
    Masses and $f_\pi$ are given in units of GeV, while $\lambda$ is expressed in 
    $\mathrm{GeV}^{-1}$}{\label{tab:basepara}}
    \begin{ruledtabular}
        \begin{tabular}{lccccccc}
        Coupling & $f_\pi$ & $g_a$ & $g_V$ & $\beta$ & $\lambda$ & $g_s$ & \\
        & 0.13025 & 0.57 & 5.8 & 0.9 & 0.56 & 0.76 & \\
        Mass & $m_D$ & $m_{D^*}$ & $m_\pi$ & $m_\eta$ & $m_\rho$ & $m_\omega$ & $m_\sigma$ \\
        & 1.867 & 2.009 & 0.137 & 0.548 & 0.775 & 0.783 & 0.600 \\
        \end{tabular}
    \end{ruledtabular}
\end{table}

\subsection{$D^{(*)}K$ interaction}

In covariant chiral perturbation theory, the leading-order (LO) $DK$ interaction 
is dominated by the Weinberg--Tomozawa term, which in the nonrelativistic limit 
can be treated as a constant potential,  
    \begin{align}
        V_{DK}(\, \vec{q} \,) = - \frac{C_W(I)}{2f_\pi^2},
    \end{align}
where $C_W(0)=2$ and $C_W(1)=0$, that is, the $I=1$ channel is noninteracting at LO. 
Performing a Fourier transform on $V_{DK}({\vec q})$ and introducing a local Gaussian regulator yields
    \begin{align}
        V_{DK}(\, r \,) = - \frac{C_W(I)}{2f_\pi^2} \frac{e^{-(r/R_c)^2}}{\pi^{3/2}R_c^3}.
    \end{align}
The next-to-leading-order (NLO) correction to the $I=0$ $DK$ S-wave interaction is 
repulsive \cite{PhysRevD.89.014026}. A repulsive core is therefore included to improve 
the description of the interaction. On this basis, the following form of $V_{DK}$ is adopted~\cite{Wu:2019vsy}:
    \begin{align}
        V_{DK}(\vec{r},R_C) = C_S \, e^{-(r/R_s)^2} + C_L \, e^{-(r/R_C)^2},
    \end{align}
where the first term accounts for the short-range repulsive core with $C_S \geq 0$ and $R_S$ fixed to $0.5$~fm, 
while the second term provides the dominant long-range attraction.

For $V_{D^*K}$, the corresponding LO and NLO potentials can be derived from the same 
covariant chiral Lagrangian \cite{PhysRevD.89.014026}:
    \begin{align}
        &V_{\mathrm{LO(NLO)}}(D^*(p_1)K(p_2) \to D^*(p_3)K(p_4)) \nonumber\\
        &=-\epsilon_3^* \cdot \epsilon_1 V_{\mathrm{LO(NLO)}}(D(p_1)K(p_2) \to D(p_3)K(p_4)).
    \end{align}
Under the infinite heavy-quark limit, $m_Q \to \infty$, this reduces to
\begin{align}
    &V_{\mathrm{LO(NLO)}}(D^*(p_1)K(p_2) \to D^*(p_3)K(p_4)) \nonumber\\
    &=V_{\mathrm{LO(NLO)}}(D(p_1)K(p_2) \to D(p_3)K(p_4)).
\end{align}
Assuming that $D_{s0}^*(2317)$ is a $DK$ bound state, the model parameters can be determined 
by reproducing its pole position. However, since the potential contains four parameters, a fit to 
a single pole is underconstrained. To obtain a more robust interaction, we also compute the 
scattering length  by solving the Lippmann--Schwinger (LS) equation and 
compare these results with lattice QCD simulations.

\section{\label{sec:level3}Three-body formalism}

\subsection{Wave function}
\begin{figure}
  \centering
  \includegraphics[width=0.5\textwidth]{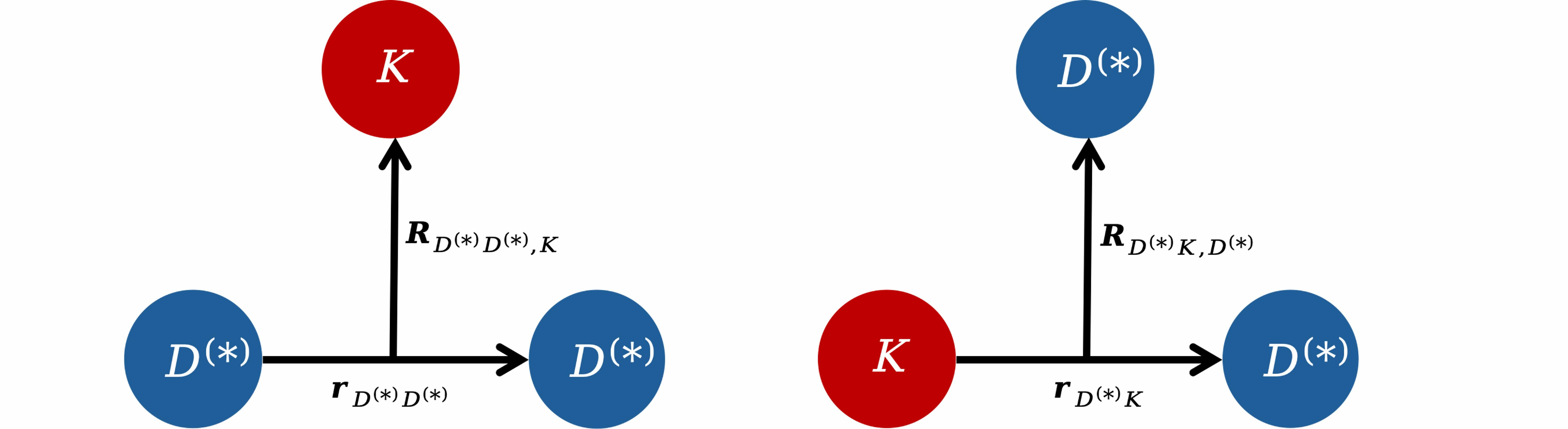}
  \caption{Two sets of Jacobi coordinates corresponding to different spatial
  configurations, where blue disks indicate the $K$ meson and the red disk
  denotes the $D^{(\ast)}$ meson. The third set of Jacobi coordinates, obtained by
  exchanging the two $D^{(\ast)}$ in panel (b), is not shown explicitly, but is
  included in the calculation.}
  \label{fig:jacobi}
\end{figure}
We consider the $D^{(*)}D^{(*)}K$ system with $I(J^P)=\frac{1}{2}(0^{-})$.
Since orbital excitations are difficult to treat accurately in GEM, we restrict
our analysis to S-wave configurations. The wave function is written as
\begin{align}
    \Psi \propto S \otimes I \otimes R
\end{align}
where $S$, $I$, and $R$ are the spin, isospin, and spatial wave functions,
respectively. Their explicit forms are given in the following subsections.

\subsection{Gaussian expansion method}

We employ the Gaussian expansion method (GEM) \cite{Hiyama:2003cu,hiyama2024} 
to solve the three-body Schr\"odinger equation:
    \begin{align}
        \hat{H} \Psi_{JM} = E \Psi_{JM},
    \end{align}
where $\hat{H}$ denotes the total Hamiltonian and is given by
    \begin{align}
        \hat{H} &= \sum_{i=1}^{3} \frac{p_i^2}{2m_i} - T_{\text{c.m.}} + \sum_{1 \leq i<j \leq 3} V(r_{ij}) \nonumber \\
                &= - \frac{\hbar^2}{2\mu_{r}} \nabla_{\vec{r}}^2
                - \frac{\hbar^2}{2\mu_{R}} \nabla_{\vec{R}}^2
                + \sum_{1 \leq i<j \leq 3} V(r_{ij})
    \end{align}    
with $T_{\text{c.m.}}$ being the center-of-mass (CM) energy. ${\vec r}$ and ${\vec R}$ represent the Jacobi coordinates. 

In GEM, the total three-body state is expanded as
    \begin{align}
        \Psi_{JM} = \sum_{\alpha,c} C_{\alpha,c} \Psi_{JM,\alpha}^c(\boldsymbol{r}_{c},\boldsymbol{R}_{c}),
    \end{align}   
where $\alpha=\{nN,lL\Lambda,tT\}$ denotes all the quantum numbers. Here $l$ and $L$ are 
the orbital-angular momentum quantum numbers for the two Jacoobi coordinates respectively,  and 
$\Lambda$ is the total orbital momentum quantum number. $t$ and $T$ denote
the isospin of two-body subsystem and the total isospin, respectively. $n$ and $N$ are the 
indices of Gaussain basis for the two Jacobi coordinates. In the present work,  due to the numerical 
calculation issue, we restrict $l = 0$ and $L = 0$, which will be explained in detail later.  
Therefore, $\Lambda$ is fixed to 0. Now, the individual basis states are defined as
    \begin{align}
        \Psi_{JM,\alpha}^c(\boldsymbol{r}_{c},\boldsymbol{R}_{c}) = 
        I_{T,t}^c \otimes \left[ S_{JM} \otimes \Phi_{lL,\Lambda}^c \right]_{JM}.
    \end{align}   
Here, $I_{T,t}^c$ and $\Phi_{lL,\Lambda}^c$ represent the isospin and spatial wave functions, respectively. 
In our calculations, we consider different isospin-coupling schemes for the three-body system, where the 
isospin wave function is given by
    \begin{align}
        &I_{T,t}^{c=1} = \left[ \left[ D_1\left(\frac{1}{2}\right) D_2\left(\frac{1}{2}\right) \right]_{1} K\left(\frac{1}{2}\right) \right]_{\frac{1}{2}}, \nonumber \\
        &I_{T,t}^{c=2} = \left[ \left[ D_2\left(\frac{1}{2}\right) K\left(\frac{1}{2}\right) \right]_{t} D_1\left(\frac{1}{2}\right) \right]_{\frac{1}{2}}, \nonumber \\
        &I_{T,t}^{c=3} = \left[ \left[ D_1\left(\frac{1}{2}\right) K\left(\frac{1}{2}\right) \right]_{t} D_2\left(\frac{1}{2}\right) \right]_{\frac{1}{2}},
    \end{align}    
with $X\left(\frac{1}{2}\right)$ denoting the isospin of particle $X$ is $\frac{1}{2}$.
Due to Bose-Einstein statistics, the isospin for the subsystem of two identity $D$ mesons can be ony 1.
In the expression, we have already set the total isospin to $T=\frac{1}{2}$. 

Considering the spin quantum numbers: $J(D)=0$ and $J(K)=0$, the spin part $S_{JM}$ of the $DDK$ system is trivial. 
For $J(D^*)=1$, the spin part $S_{JM}$ of the $D^*D^*K$ system can be expressed as 
    \begin{align}
        S_{JM} =S_{00}= \frac{1}{\sqrt{3}}\ket{1\,-1} + \frac{1}{\sqrt{3}}\ket{-1\,1} - \frac{1}{\sqrt{3}}\ket{0\,0},
    \end{align}
 where the coefficients in front of the states are Clebsch-Gordan (CG) coefficients. 
    
In the CMS frame, we rewrite the Schr\"odinger equation in terms of Jacobi coordinates and expand the 
coordinate-space wave function using two sets of Gaussian basis functions, $\phi_{nlm}^G(\boldsymbol{r}_c)$ and $\psi_{NLM}^G(\boldsymbol{R}_c)$, corresponding two Jacobi coordinates:
    \begin{align}
        &\Phi_{lL,\Lambda}^c(\boldsymbol{r}_{c},\boldsymbol{R}_{c}) = 
         \left[\phi_{nlm}^G(\boldsymbol{r}_c) \psi_{NLM}^G(\boldsymbol{R}_c)\right]_\Lambda , \nonumber \\
        &\phi_{nlm}^G(\boldsymbol{r}) = N_{nl} r^l e^{-\nu_n r^2} Y_{lm}(\hat{\boldsymbol{r}}), \nonumber \\
        &\psi_{NLM}^G(\boldsymbol{R}) = N_{NL} R^L e^{-\lambda_{N} R^2} Y_{LM}(\hat{\boldsymbol{R}}), \nonumber \\
        &N_{nl} = \left( \frac{2^{l+2}(2\nu_n)^{l+\frac{3}{2}}}{\sqrt{\pi}(2l + 1)!!} \right)^{\frac{1}{2}},
    \end{align}
where $Y_{LM}$ is the spherical harmonic function. Three sets of Jacobi coordinates are considered in our system, as two of them illustrated in Fig.~\ref{fig:jacobi}.  It is clear that the basis involves powers of coordinates for nonzero $l$ and $L$, which introduces 
considerable numerical challenges in the computation of matrix elements. Since the contribution 
from higher partial waves is expected to be much smaller than that of the S-wave, in this work, 
we only consider the ground state of the three-body system.

To maximize the numerical accuracy, the Gaussian parameters are chosen in geometric progression:
    \begin{align}
     &\lambda_N = 1/R_N^2, \quad R_N = R_{\text{min}} A^{N-1} \quad (N=1, 2, \dots, N_{\text{max}}), 
     \nonumber \\
        &\nu_n = 1/r_n^2, \quad r_n = r_{\text{min}} a^{n-1} \quad (n=1, 2, \dots, n_{\text{max}}).
           \end{align}
In the present work, the values of the size parameters are taken as
    \begin{align}
        r_1=0.1\,\mathrm{fm}, \qquad r_{\mathrm{max}}=10\,\mathrm{fm}, \qquad n_{\mathrm{max}}=12.
    \end{align}
    
To account for the coupled channels $DDK$ and $D^*D^*K$, we solve the coupled-channel Schr\"odinger equation:
    \begin{align}
        \left[ \left( -\frac{\hbar^2}{2\mu_r} \nabla_{\vec{r}}^2 - \frac{\hbar^2}{2\mu_{R}} \nabla_{\vec{R}}^2 \right) \delta_{ij} + V_{ij} \right] \psi_j = E \psi_i,
    \end{align}
where the potential matrix $V$ is given by
    \begin{align}
        V=
        \begin{pmatrix}
            V_{DDK\to DDK} & V_{DDK \to D^*D^*K} \\
            V_{D^*D^* K \to DDK} & V_{D^*D^*K \to D^*D^*K}
        \end{pmatrix}.
    \end{align}

\subsection{Complex scaling method}
The GEM provides an efficient real-space expansion for locating bound states, 
but it is not directly applicable to resonant states with complex energies. To address this, we employ the 
complex scaling method (CSM). In this approach, the Schr\"odinger equation is analytically continued via 
a coordinate rotation in the complex plane,
\begin{align}
&\boldsymbol{r} \to \boldsymbol{r}e^{i\theta}, \qquad \boldsymbol{p} \to \boldsymbol{p}e^{-i\theta}, \nonumber\\
&\hat{H}(\theta) = \sum_{k=1}^{3} \frac{p_k^2}{2m_k}  e^{-2i\theta}- T_{\text{c.m.}} + \sum_{1\leq k<l \leq 3} V(r_{kl} e^{i\theta}).
\end{align}
The complex-scaled Hamiltonian is then diagonalized within the same Gaussian basis used in the GEM. 
The resulting eigenvalue spectrum can be visualized in the complex-energy plane: bound states appear 
on the negative real axis, while resonance poles are isolated from the rotated continuum branches. 
Most eigenvalues lie along rays emanating from thresholds and correspond to scattering states.

The root-mean-square (rms) radius provides a valuable observable for characterizing the internal 
structure of hadronic bound states, and is often used to distinguish compact multiquark configurations 
from extended molecular-type systems. It is defined as~\cite{Papadimitriou2016,Berggren1996}
\begin{align}
r_{ij}^{\mathrm{rms}} \equiv \sqrt{\frac{\bra{\Psi^{IJ}}r_{ij}^2\ket{\Psi^{IJ}}}{\braket{\Psi^{IJ}|\Psi^{IJ}}}}.
\end{align}

\section{\label{sec:level4}Results and discussion}

\subsection{Two-body $DK$ and $D^*K$}

For concreteness, we fix $R_S = 0.5$ fm and scan the remaining model parameters. 
The long-range coupling constant $C_L$ is fixed by reproducing the pole position of $D_{s0}^*(2317)$. 
Once this condition is satisfied, the $D_{s1}(2460)$ state is reproduced approximately in the corresponding 
$D^*K$ channel, as expected by the heavy-quark symmetry. For each parameter set, we solve the Lippmann–Schwinger (LS) equation for 
the two-body $DK$ system~\cite{LippmannSchwinger1950,Koike1992,Petschauer:2020}:
        \begin{align}
            \boldsymbol{T} = \boldsymbol{V} + \boldsymbol{VGT},
        \end{align}
whose partial-wave representation is given by
        \begin{align}
            T_\ell(p,q) &= V_\ell(p,q) \nonumber\\
            & + \quad \frac{\mu}{4\pi^3} \int_0^\infty dk \, k^2 V_\ell(p,k)G(k)T_\ell(k,q),
        \end{align}
where $G(k) = 1/(E - \frac{k^2}{2\mu} + i\epsilon)$ denotes the two-body propagator, and the 
partial-wave potential $V_\ell(p,q)$ is defined as
        \begin{equation}
            V_\ell(p,q) = 2\pi \int_{-1}^{1}dx\, P_\ell(x) \int d^3r\, V(r)\, e^{-i(\boldsymbol{p}-\boldsymbol{q})\cdot \boldsymbol{r}},
        \end{equation}
with $x = \hat{\boldsymbol{p}} \cdot \hat{\boldsymbol{q}}$ being the dot product of the momentum unit vectors.

For the partial-wave scattering amplitude $f_\ell$, scattering theory yields \cite{Bethe1949}
        \begin{equation}
            \frac{f_\ell(k)}{k} = -\frac{\mu}{8\pi^2} T_\ell(k,k;E) = \left( -a^{-1} + \frac{1}{2}r k^2 - ik \right)^{-1},
        \end{equation}
where $E = \frac{k^2}{2\mu}$ is the two-body energy, and $a$ and $r$ are the scattering length and effective range, respectively.

\begin{figure*}
    \includegraphics[width=\linewidth]{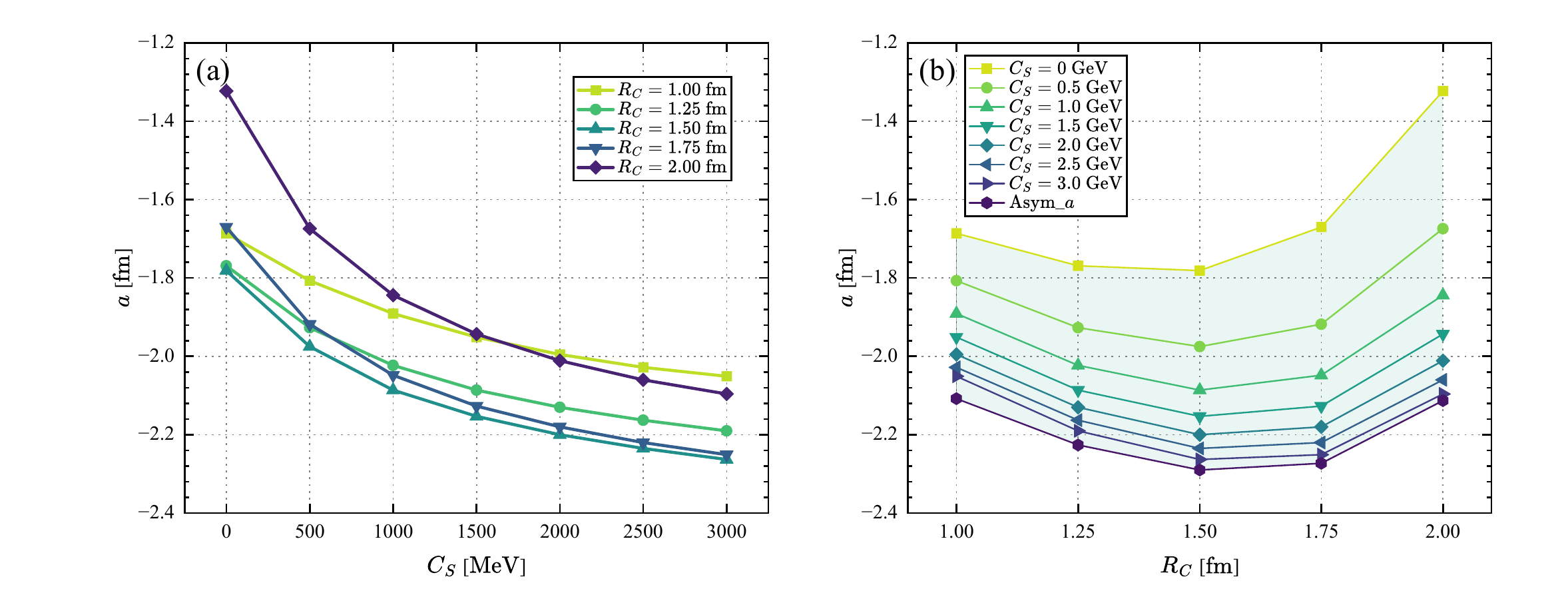}
    \caption{$(a)$ shows the scattering lengths for different $R_C$ values versus 
    coupling constant $C_S$, and $(b)$ shows the scattering lengths for different 
    coupling constants $C_S$ versus $R_C$. Asym\_a represents the asymptotic scattering lengths
    of different $R_C$, the shaded region between curves represents 
    the allowed range of scattering lengths}
    \label{fig:a}
\end{figure*}

We calculate the scattering length for $R_C = 1.0 - 2.0$ fm and $C_S = 0 - 3000$ MeV, with the corresponding scattering lengths shown in Fig.~\ref{fig:a}. We also examined the case with $R_S = 0.5\,\mathrm{fm}$ and $R_C = 3.0$ fm, which yields scattering lengths smaller than $-15$~fm and is therefore not adopted in the main analysis. As shown in Fig.~\ref{fig:a}(a), for each fixed $R_C$, the scattering length decreases monotonically with increasing $C_S$ and gradually approaches an asymptotic value. This behavior indicates that the low-energy $DK$ observables become less sensitive to the short-range repulsive core once $C_S$ is sufficiently large. By fitting these curves, we extract the asymptotic values displayed in Fig.\ref{fig:a}(b). It can be further seen that the fitted asymptotic scattering length exhibits a nonmonotonic dependence on $R_C$, reaching its most negative value around $R_C \sim 1.5$ fm. This behavior suggests that the long-range structure of the $DK$ interaction plays a nontrivial role in determining the low-energy scattering properties, which must therefore be constrained before proceeding to the three-body calculation.

According to recent lattice QCD simulations \cite{Lattice_scat_length}, the predicted S-wave scattering
 length for the $DK$ system at physical meson masses is $a_{DK} = -1.87\,\mathrm{fm}$ (central value), 
 while an alternative central value of $a_{DK} = -1.51\,\mathrm{fm}$ is also reported. The sizable discrepancy
 between these two values arise from systematic error. Using the exponential fits described above, we extract the 
$C_S$ values corresponding to these lattice scattering lengths, and then determine $C_L$ by reproducing the
 $D_{s0}^*(2317)$ pole. The resulting parameter sets are listed in Table \ref{tab:centralvalue}, which will be primarily used in our subsequent three-body calculations.

\begin{table}[h]
\caption{Scattering lengths~\cite{Lattice_scat_length} and their corresponding $DK$ parameters.}
\label{tab:centralvalue}
\begin{ruledtabular}
\begin{tabular}{cccc}
$a$ (fm) & $R_C$ (fm) & $C_S$ (MeV) & $C_L$ (MeV) \\
\colrule
-1.87 & 1.00 & 862.3  & -535.3 \\
-1.87 & 1.25 & 304.9  & -298.1 \\
-1.87 & 1.50 & 206.3  & -225.0 \\
-1.87 & 1.75 & 404.1  & -202.9 \\
-1.87 & 2.00 & 1089.0 & -197.3 \\
-1.51 & 2.00 & 243.2  & -165.8 \\
\end{tabular}
\end{ruledtabular}
\end{table}

\subsection{$DDK$ spectrum}
We now turn to the three-body $DDK$ system with $I(J^P) = \frac{1}{2}(0^-)$ within a coupled-channel framework that incorporates both  $DDK$ and $D^*D^*K$ configurations. The $DK$ interaction is modeled by a two-range Gaussian potential, whose parameters are constrained by the pole position of $D_{s0}^*(2317)$ and the $DK$ scattering length from lattice QCD calculations. Since there are three free parameters but only two constraint conditions from experiment and lattice simulations, we select two representative parameter sets based on the $DK$ scattering length predicted by lattice QCD.

For the $D^{(*)}D^{(*)}$ interaction within the OBE model, the coupling constant for $\sigma$-exchange is constrained by the $Z_c(3900)$ pole, while those for vector meson exchange are constrained by the poles of $X(3872)$ and $T_{cc}^+$. Specifically, the poles of $T_{cc}^+$ and $X(3872)$ are fixed to $-4~\mathrm{MeV}$ relative to their respective thresholds, whereas the pole associated with $Z_c(3900)$ is varied within the range of $-5$ to $-35~\mathrm{MeV}$.

\begin{figure*}[htp]
    \includegraphics[width=1\linewidth]{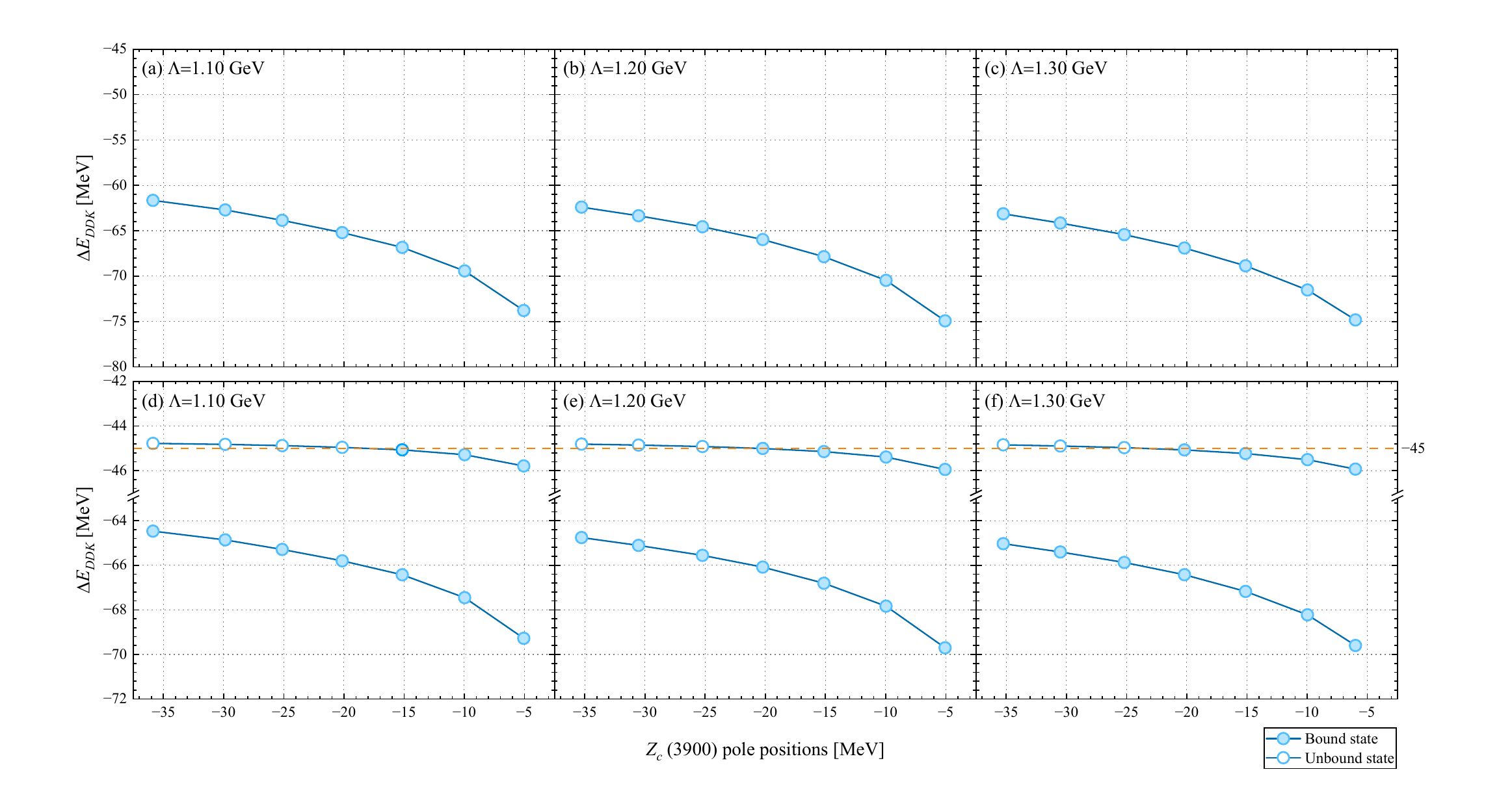}
    \caption{The cutoff dependence of the $DDK$ bound states. The open (filled) 
    circles represent unbound (bound) states. 
    The orange line indicates the particle-dimer($D-DK$) threshold. 
    (a)--(c): $R_C$=1.0\,fm, $C_S$=0, $C_L$=320.1\,MeV. 
    (d)--(e): $R_C$=2.0\,fm, $C_S$=500\,MeV, $C_L$=178.4\,MeV.}
    \label{fig:bound}
\end{figure*}

The resulting three-body spectra are displayed in Fig.~\ref{fig:bound}. For both $DK$ parameter sets, the $DDK$ system supports bound states over a wide range of $DD$ interaction strengths. As shown in Fig.~\ref{fig:bound}, the binding energies exhibit only a mild dependence on the momentum cutoff in the range $\Lambda = 1.1 - 1.3$ GeV, indicating that the results are reasonably stable against short-range regularization of the $D^{(*)}D^{(*)}$ interaction. This weak cutoff dependence suggests that the existence of bound-state solutions is not a numerical artifact of a specific regulator choice, but rather reflects a robust feature of the underlying two- and three-body dynamics within the present framework.

Depending on the $DK$ parameter choice, two scenarios emerge.
For the first scenario, which typically corresponds to a smaller $R_C$, 
only a single bound state is found. Its binding energy lies 
in the range of approximately $-60$ to $-75$ MeV relative to the three-body 
threshold and changes smoothly with the input $Z_c(3900)$ pole position.
This is consistent with the result of \cite{Wu:2019vsy}, which is around $-70$~MeV and 
varies slightly as using different potential parameters. 
In contrast, for the second scenario with a larger $R_C$, two bound states 
emerge in most cases: a deeply bound state 
and a shallow state located close to the particle-dimer ($D-DK$) threshold. This 
demonstrates that the long-range structure of the $DK$ interaction 
plays a crucial role in generating additional three-body solutions.
We further analyze the role of coupled-channel dynamics by examining 
the probability of the $D^*D^*K$ component in the total wave function. 
It is found that, in all cases considered, the $D^*D^*K$ fraction is 
extremely small, remaining below $0.1\%$. This indicates that the 
contribution from the $D^*D^*K$ channel is negligible, and the structure 
of the system is overwhelmingly dominated by the $DDK$ component.
This justifies, \textit{a posteriori}, the use of a simplified model space 
when discussing the dominant structure and spatial properties of the system.

\begin{figure*}[htp]
    \includegraphics[width=1\linewidth]{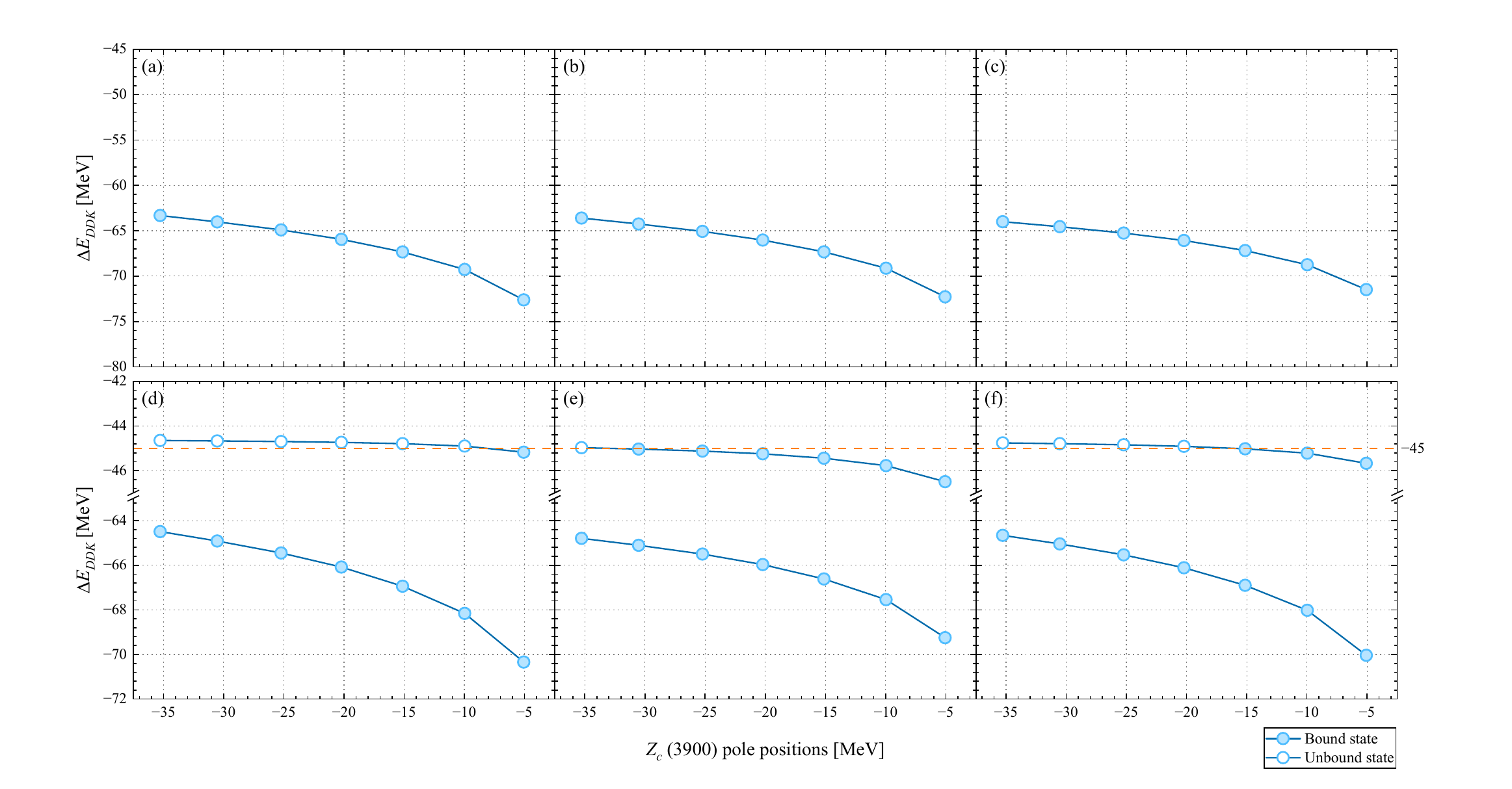}
    \caption{The $DDK$ bound states vary with the $Z_c(3900)$ virtual state pole position, 
    given that the $DK$ interaction is determined by fitting the lattice-calculated scattering 
    length of $DK$. The momentum cutoff for the $DD$ interaction is fixed to $\Lambda = 1.2$~GeV. Panels (a)--(e) correspond to $a_{DK} = -1.87$ fm with $R_C = 1.00$, $1.25$, $1.50$, $1.75$, and $2.00$ fm, respectively; panel (f) corresponds to $a_{DK} = -1.51$ fm with $R_C = 2.00$ fm.}
    \label{fig:central_bound}
\end{figure*}

Fixing the cutoff at $\Lambda=1.2$\,GeV, we adopt the $DK$ parameters
filtered according to the scattering lengths presented in
Table \ref{tab:centralvalue}. The corresponding binding energies are plotted in Fig.\ref{fig:central_bound}.
As shown in the figure, once the $DK$ interaction is constrained by the lattice-derived 
scattering length of $DK$, the binding energies are confined to a narrower range. 
Meanwhile, the dependence on $R_C$ is pretty mild. 

\subsection{Internal structure of $DDK$}
\begin{figure*}[htp]
    \includegraphics[width=\linewidth]{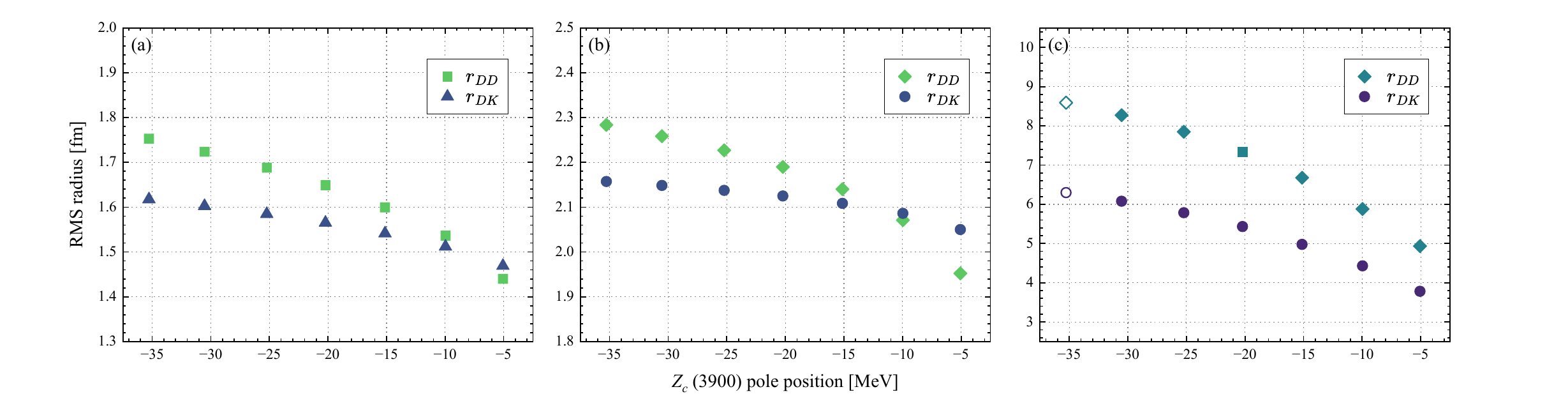}
    \caption{(Color online) The root-mean-square (rms) radii of the two-body subsystems $DD$ and $DK$ in the three-body $DDK$ system, where the $DK$ interaction is determined by fitting the lattice-calculated scattering length of the $DK$ system, 
    $a_{DK} = -1.87$~fm. Panels (a)--(c) are defined as follows: (a) corresponds to $R_C = 1.00$ fm; 
    (b) shows the deep bound state for $R_C = 2.00$ fm; (c) shows the shallow bound state for 
    $R_C = 2.00$ fm. Moreover, open shapes denote unbound states, while filled shapes denote bound states.}
    \label{fig:central_rms}
\end{figure*}

To clarify the internal structure of the bound states, we analyze the 
root-mean-square (rms) radii shown in Fig~.\ref{fig:central_rms}(a-b). We find
    \begin{equation}
    r_{DD} \sim r_{DK} \sim 1 - 2~\mathrm{fm}.
    \end{equation}
This indicates that all three constituents are within a typical hadronic distance, with no clear separation of scales among the interparticle distances. The wave function is therefore dominated by genuine three-body correlations rather than a two-body cluster structure. In particular, neither a $(DD)+K$ nor a $(DK)+D$ configuration is favored. We thus identify these states as a compact three-body bound state, whose binding is collectively generated by the interplay of all pairwise interactions.

In Fig.~\ref{fig:central_bound}(d)--(f), an additional shallow bound state emerges 
close to the threshold. A key feature of this state is its large spatial extent, as illustrated in Fig.~\ref{fig:central_rms}(c):
    \begin{equation}
    r_{DD} \sim 5 - 9~\mathrm{fm}, \quad
    r_{DK} \sim 4 - 7~\mathrm{fm}.
    \end{equation}
All interparticle distances are significantly larger than the typical interaction range ($\sim 1$ fm), 
and importantly,
    \begin{equation}
    r_{DD} \sim r_{DK}.
    \end{equation}
These features clearly distinguish the shallow bound state from a conventional hadron. 
For a $(DK)+D$ configuration, one would expect a pronounced scale separation: $r_{DK} \ll r_{DD}$, 
whereas for a $(DD)+K$ structure, one would expect $r_{DD} \ll r_{DK}$. Neither pattern is observed 
in our results. Instead, the comparable and large values of all rms radii indicate that the three particles 
are weakly bound and spatially extended, without forming a dominant two-body subcluster. This is a characteristic signature of a three-body ``halo" state~\cite{halo}, in which the wave function has substantial support at distances far beyond the range of the pairwise interactions. The halo interpretation is further supported by the trend that the spatial extension increases as the binding energy approaches the threshold, consistent with the expected scaling behavior
    \begin{equation}
    r \propto \frac{1}{\sqrt{E_B}}.
    \end{equation}

\subsection{$D^*D^*K$ system}

\begin{figure*}
    \includegraphics[width=1\linewidth]{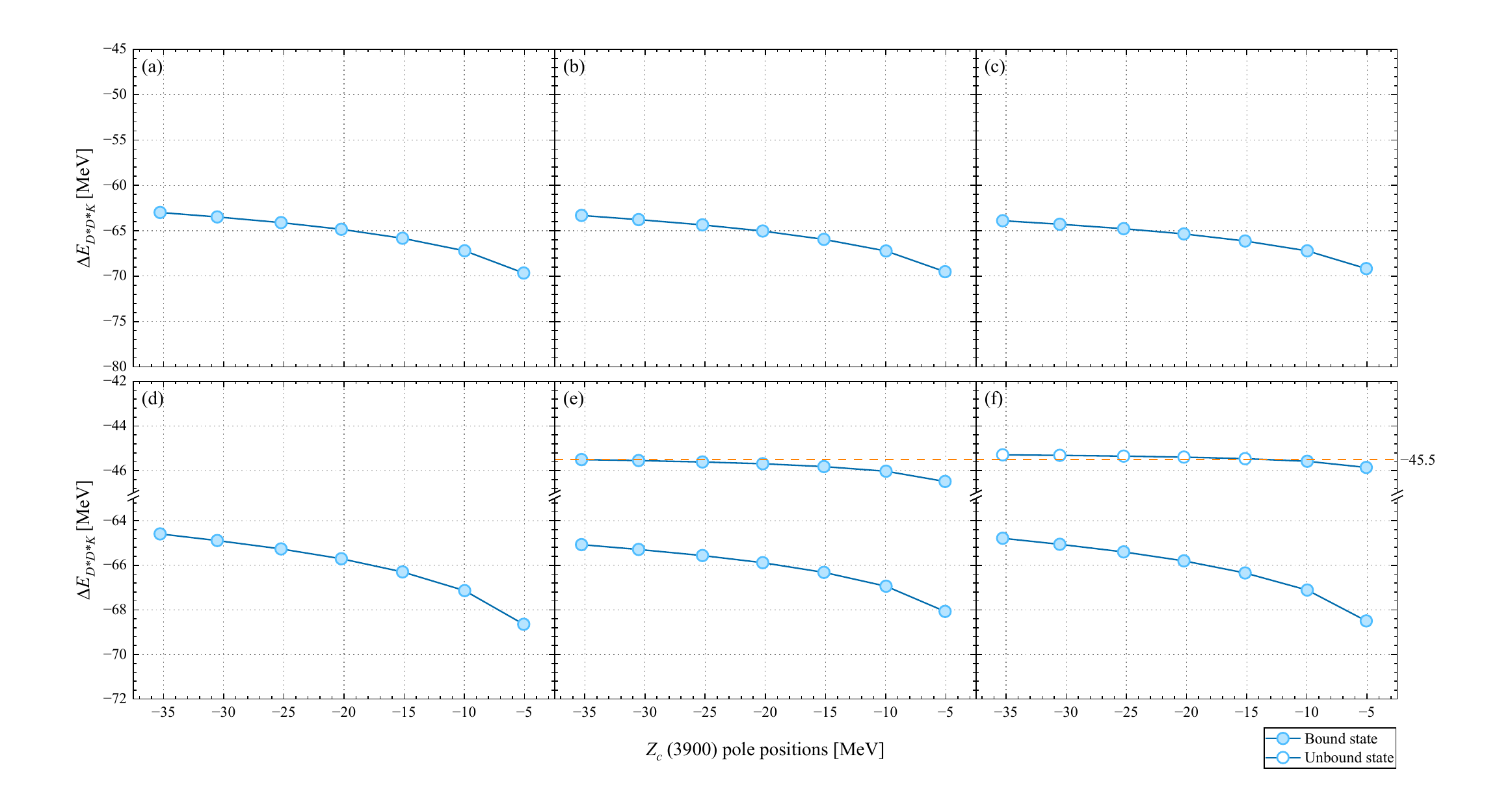}
    \caption{The three-body $D^*D^*K$ bound states versus the $Z_c(3900)$ virtual state
    pole position. The parameter settings
    of (a)--(f) are the same as those in Fig.\ref{fig:central_bound}.}
    \label{fig:central_bound2}
\end{figure*}

\begin{figure*}
    \includegraphics[width=1\linewidth]{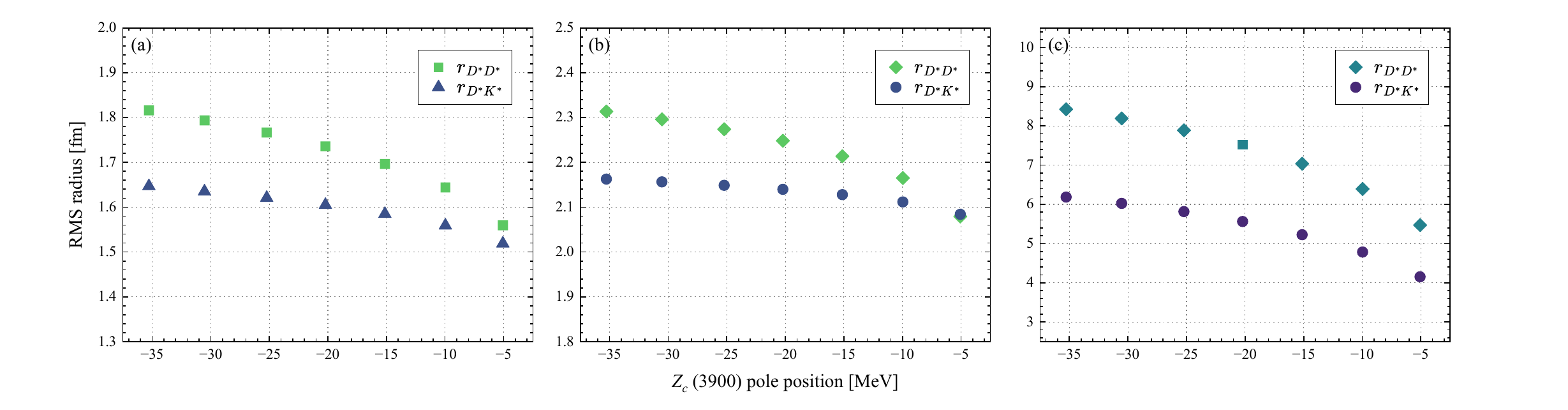}
    \caption{The rms radii of the two-body subsystems in $D^*D^*K$ system versus the $Z_c(3900)$ virtual-state pole position. The $D^*K$ interaction is determined by fitting the lattice-calculated 
    scattering length of $DK$, $a_{DK}=-1.87$~fm. Panels (a)--(c) are defined as follows: 
    (a) corresponds to $R_C = 1.00$ fm; 
    (b) shows the deep bound state for $R_C = 2.00$ fm; 
    (c) shows the shallow bound state for $R_C = 2.00$ fm.}
    \label{fig:central_rms2}
\end{figure*}

We also conduct a dedicated investigation of the $D^*D^*K$ system, independent 
of the previously analyzed $DDK$ system, to comprehensively understand its bound 
state characteristics and dynamic behavior. The numerical results obtained using the 
parameter set listed in Table~\ref{tab:centralvalue} are presented in 
Figs.~\ref{fig:central_bound2} and \ref{fig:central_rms2}, which clearly illustrate the 
binding energy and spatial properties for the subsystems $D^*D^*$ and $D^*K$. The bound 
states of the $D^*D^*K$ system exhibit a behavioral pattern very similar to  that of the 
$DDK$ system: the deeply bound state lies in the range of approximately $-60$ to $-70$ 
MeV, while the shallow bound state is positioned just below the particle-dimer ($D^*-D^*K$) threshold. 
Regarding the internal 
structure of the three-body $D^*D^*K$ system, the deeply bound state is identified 
as a compact three-body bound state, with the interparticle distances $r_{D^*D^*}$ 
and $r_{D^*K}$ both in the range of $1 - 2$ fm, indicating a tight and compact spatial 
configuration. In contrast, the shallow bound state is identified as a three-body halo 
state, which is characterized by relatively large rms radii for both
$r_{D^*D^*}$ and $r_{D^*K}$, reflecting a loose and extended spatial distribution.

\begin{figure*}
    \includegraphics[width=0.8\linewidth]{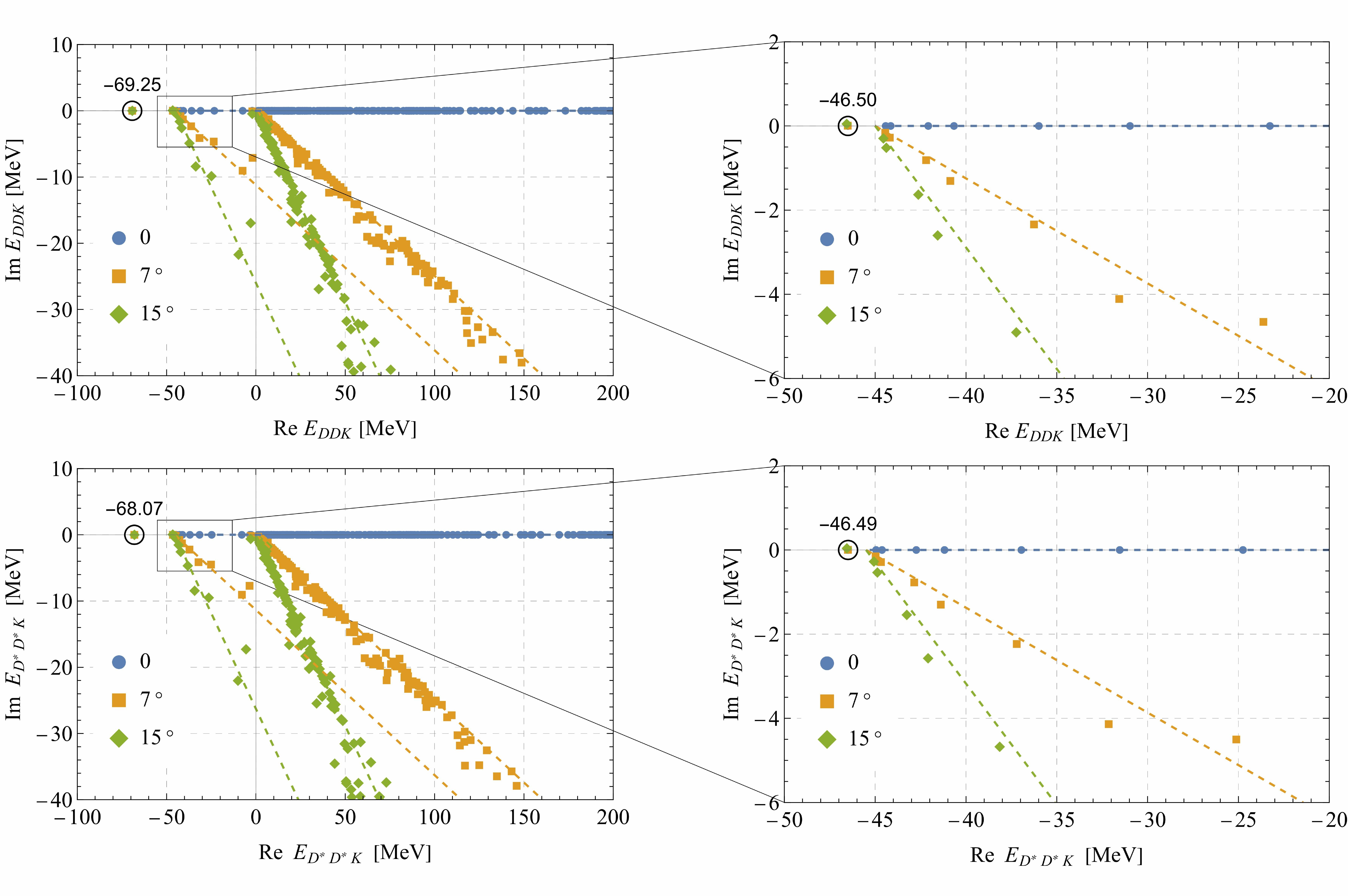}
    \caption{Complex-energy results of $DDK$(upper panel) and $D^*D^*K$(lower panel) as 
    the $DK$ interaction is determined from the lattice-calculated scattering length of 
    $DK$, $a_{DK}=-1.87$~fm and $R_C=2.00$~fm. Bound states are marked by circles.}
    \label{fig:csm}
\end{figure*}

To further verify the nature of the bound states, we perform a detailed CSM analysis, 
with Fig.~\ref{fig:csm} serving as a typical example. The CSM results clearly show that 
there are no resonance poles in the complex energy plane for either the $DDK$ or $D^*D^*K$ 
system. Specifically, the bound states of both systems appear as discrete eigenvalues 
located on the negative real axis of the complex energy plane, and these eigenvalues 
remain invariant under the complex rotation operation, consistent with the fundamental 
characteristics of genuine bound states. Meanwhile, the continuum states of the systems 
rotate in the complex energy plane as expected, which is a typical feature of non-resonant 
continuum states. Additionally, no states that remain invariant under the complex rotation 
operation are found in the third and fourth quadrants of the complex energy plane, further 
confirming the absence of resonant states. These results collectively demonstrate 
that the bound states identified in our work are genuine three-body bound states, rather 
than resonant states or numerical artifacts, and the framework we adopted is reliable 
and effective for analyzing the three-body systems.

\section{\label{sec:level5}Summary}

Although several investigations have been conducted so far, the structure of 
the $DDK$ system remains incompletely understood. In this work, we investigate 
the $I(J^P) = \frac{1}{2}(0^-)$ $DDK$ system within a coupled-channel framework 
that includes both the $DDK$ and $D^*D^*K$ channels. In particular, the $DD$ 
interaction is constrained by the experimental data associated with 
$Z_c(3900)$, $X(3872)$, and $T_{cc}^+$, while the $DK$ interaction is
fixed by the pole position of $D_{s0}^*(2317)$ together with $DK$ scattering length 
from lattice-QCD calculations. Within this constrained theoretical 
setup, we solve the three-body problem using GEM and searched for possible 
resonant states via the CSM.

We find that the $DDK$ system supports a deeply bound state over a broad 
range of parameter choices, with a typical binding energy of approximately 
$70$ MeV below the three-body threshold. The binding energy of this state 
exhibits only weak dependence on the momentum cutoff in the 
range $\Lambda = 1.1$–$1.3$ GeV, indicating that this state is a robust 
prediction within the present framework. In contrast, an additional shallow 
bound state emerges only for a subset of the parameter space, specifically, 
when the $DK$ interaction contains a short-range repulsive core alongside 
a longer-range attractive component, and may disappear into the continuum 
as the $DD$ interaction strength is varied. Moreover, an analogous study 
was applied to the $D^*D^*K$ system, in which we found that its bound-state 
behavior is similar to that of the $DDK$ system.

Analysis of the rms radii indicates that the deeply bound state possess a 
compact three-body structure, whereas the shallow state is a candidate of a  
three-body halo state. This halo state is characterized by large and comparable 
values of $r_{D^{(*)}D^{(*)}}$ and $r_{D^{(*)}K}$, as well as the absence of 
a dominant two-body subcluster. We further find that the $D^*D^*K$ admixture 
is negligible, implying that coupled-channel effect from $D^*D^*K$ plays only a minor role 
in the formation mechanism of these states. Finally, the complex scaling analysis 
reveals no additional three-body resonance poles in the parameter region 
explored in this work. The current study is expected to provide valuable 
information for understanding the rich structures of the three-body $DDK$ 
and $D^*D^*K$ systems, as well as for the future experimental search 
for such exotic hadronic molecules at facilities such as BESIII, LHCb and Belle II.

\begin{acknowledgements}
We thank Lu Meng for useful suggestions. 
This work is supported by the Special Funds for Theoretical Physics under the National Natural Science 
Foundation of China (Grant No. 12547105) and the National Natural Science Foundation of China (Grant No. 12575153). 
\end{acknowledgements}

\section*{Data Availability}
The data that support the findings of this study are openly available in Zenodo repository at http://doi.org/10.5072/zenodo.484495.
\bibliographystyle{apsrev4-2}
\bibliography{ref}

\end{document}